\documentclass[letter,traditabstract]{aa} % for the abstract without structuration 
\usepackage{graphicx}
\usepackage{subfigure}
\usepackage{txfonts}
\usepackage{appendix}
\newcommand{\hen}{Hen$\,$3-1379}
\newcommand{\wra}{Wray15-751}
\newcommand{\irc}{IRC$+$10420}
\begin{document}
   \title{\textit{A massive parsec-scale dust ring nebula around 
         the yellow hypergiant \hen }
         }

   \author{D. Hutsem\'{e}kers\inst{1}\fnmsep\thanks{Senior Research Associate F.R.S.-FNRS}
          \and N.L.J. Cox\inst{2} 
          \and C. Vamvatira-Nakou\inst{1}
          }

   \institute{Institut d'Astrophysique et de G\'{e}ophysique, Universit\'{e} de Li\`ege, 
              All\'{e}e du 6 ao\^ut, 17 - B5c, B-4000 Li\`ege, Belgium \\
              \email{hutsemekers@astro.ulg.ac.be}
           \and Instituut voor Sterrenkunde, KU Leuven, 
                Celestijnenlaan 200D, B-3001 Leuven, Belgium 
           }

   \date{Received , 2013; accepted , 2013}

% \abstract{}{}{}{}{} 
% 5 {} token are mandatory
 
  \abstract{
On the basis of far-infrared images obtained by the {\it Herschel}
Space Observatory, we report the discovery of a large and massive dust
shell around the yellow hypergiant \hen . The nebula appears as a
detached ring of 1 pc diameter which contains 0.17 $M_{\odot}$ of
dust. We estimate the total gas mass to be 7 $M_{\odot}$, ejected some
$1.6\times 10^{4}$ years ago. The ring nebula is very similar to
nebulae found around luminous blue variables (LBVs) except it is not
photoionized.  We argued that \hen\ is in a pre-LBV stage, providing
direct evidence that massive LBV ring nebulae can be ejected during
the red supergiant phase.
}


   \authorrunning{D. Hutsem\'ekers  et al.}
   \titlerunning{A  massive parsec-scale dust ring nebula around 
                 the yellow hypergiant \hen }

   \maketitle

%
%________________________________________________________________

\section{Introduction}
\label{sec:intro}

According to current evolutionary models of massive stars (e.g.,
Maeder and Meynet \cite{mae10}), O-type stars evolve into Wolf-Rayet
stars by losing a significant fraction of their initial mass.
Luminous blue variable stars (LBVs) represent a short (a few
$10^{4}$~yr) unstable phase in the upper part of the
Hertzsprung-Russell (HR) diagram.  Their luminosities are typically
5.3 $ < \log L / L_{\odot} < $ 6.3 and their spectral types between O9
and A (Humphreys and Davidson \cite{hum94}). LBVs are often surrounded
by massive dusty nebulae (Hutsem\'{e}kers \cite{hut94}; Nota et
al. \cite{not95}) which reveal episodes of extreme mass-loss. LBVs
likely represent an intermediate stage between main-sequence O-type
and Wolf-Rayet stars, for stars with initial mass higher than
$\sim$30~$M_{\odot}$.

It is not clear at which evolutionary stage the massive
($\geq$1~$M_{\odot}$) nebulae observed around LBVs are ejected, and
what is the physical mechanism responsible for their ejection.  Based
on a dust composition found to be very similar to that of red
supergiants (RSGs), Waters et al. (\cite{wat97}) and Voors et
al. (\cite{voo00}) argued that LBV nebulae were ejected when the stars
were RSGs, although no RSG has been observed with $\log L / L_{\odot}
\geq$ 5.8. By comparing the N/O abundances in the ejected nebulae to
the predicted surface composition of massive stars during various
phases of their evolution, Lamers et al. (\cite{lam01}) concluded that
LBV nebulae are ejected during the blue supergiant phase and that the
stars have not gone through a RSG phase. But Lamers et al.
(\cite{lam01}) only considered a sample of very luminous LBVs, all
with $\log L / L_{\odot} > $ 5.8.  On the other hand, the
determination of C, N, O abundances in the nebula around the less
luminous LBV \wra , based on far-infrared spectroscopy with the {\it
Herschel} space observatory, led to the conclusion that the nebula was
ejected during a RSG phase (Vamvatira-Nakou et al., in prep.).

Here we report the discovery of a massive parsec-scale dust nebula
around the yellow hypergiant \hen\ which has not yet reached the LBV
stage, providing direct evidence for the ejection of LBV-type nebulae
during the RSG phase.

\hen\ (= IRAS 17163-3907) is an emission-line star discovered by
Henize (\cite{hen76}). It has been studied in detail by Le~Bertre et
al. (\cite{leb89,leb93}). In particular, \hen\ was found irregularly
variable, both photometrically and spectroscopically, with strong
infrared emission due to dust, mostly silicate. These authors noticed
the strong spectral similarity with the LBV HR~CAR, but classified
\hen\ as a post-AGB star due to its low luminosity.

From a detailed study of foreground insterstellar absorption, Lagadec
et al. (\cite{lag11}) revised the distance to \hen .  They found the
star located at $\sim$4 kpc, then with a luminosity $\log L /
L_{\odot} \simeq $ 5.7, comparable to that of the LBV \wra , and a
late B or early A spectral type, characteristic of yellow hypergiants
(YHGs, e.g., Oudmaijer et al. \cite{oud09}).  Mid-infrared imaging
($\sim$10 $\mu$m) revealed the presence of two concentric dust shells
around the star, with radii 0$\farcs$6 and 1$\farcs$5, indicative of
mass-loss enhancement. By modeling the mid- to far-infrared emission,
they also predict the existence of a larger and colder dust shell.

\section{A large dust nebula around \hen }
\label{sec:imaging}

\begin{figure}[t]
\resizebox{\hsize}{!}{\includegraphics*{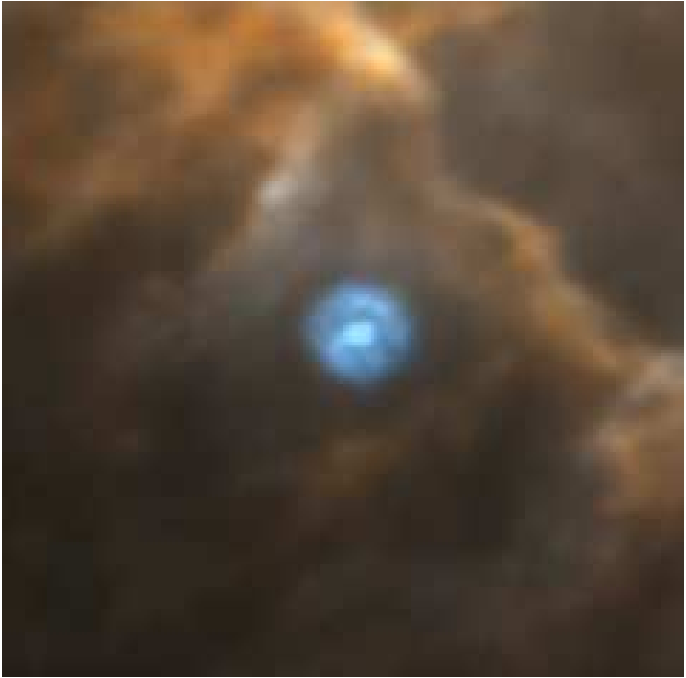}}
\caption{A two-color image of \hen\ and its environment, from {\it
Herschel} PACS observations at 70 $\mu$m (blue) and 160 $\mu$m
(red). The field is 8\arcmin $\times$ 8\arcmin\ with a the pixel size
of 2\arcsec . North is up and East to the left. The dust nebula around
\hen, roughly 50\arcsec\ in diameter, pops up over the colder
interstellar medium.}
\label{rgb}
\end{figure}

\begin{figure}[t]
\resizebox{\hsize}{!}{\includegraphics*{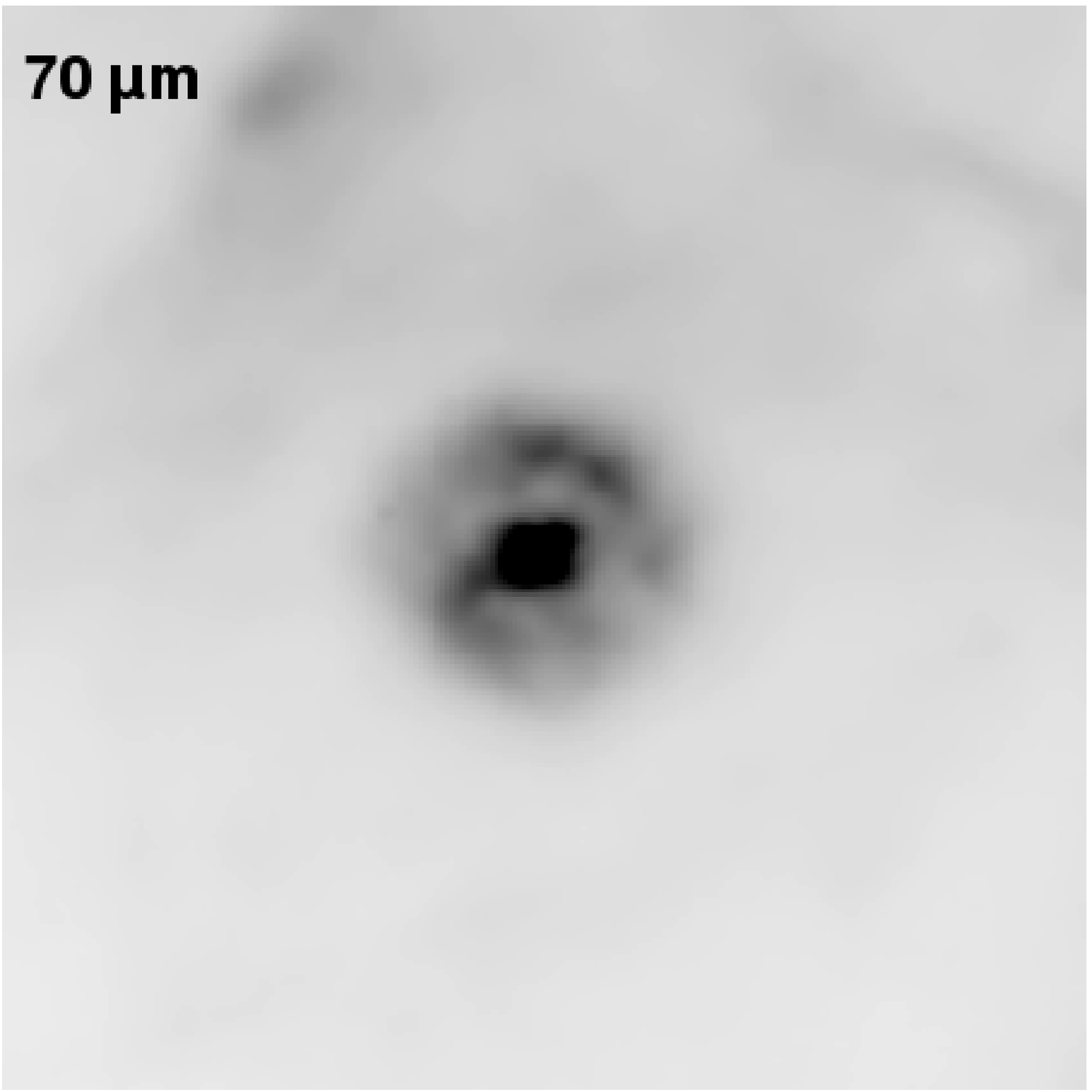}\includegraphics*{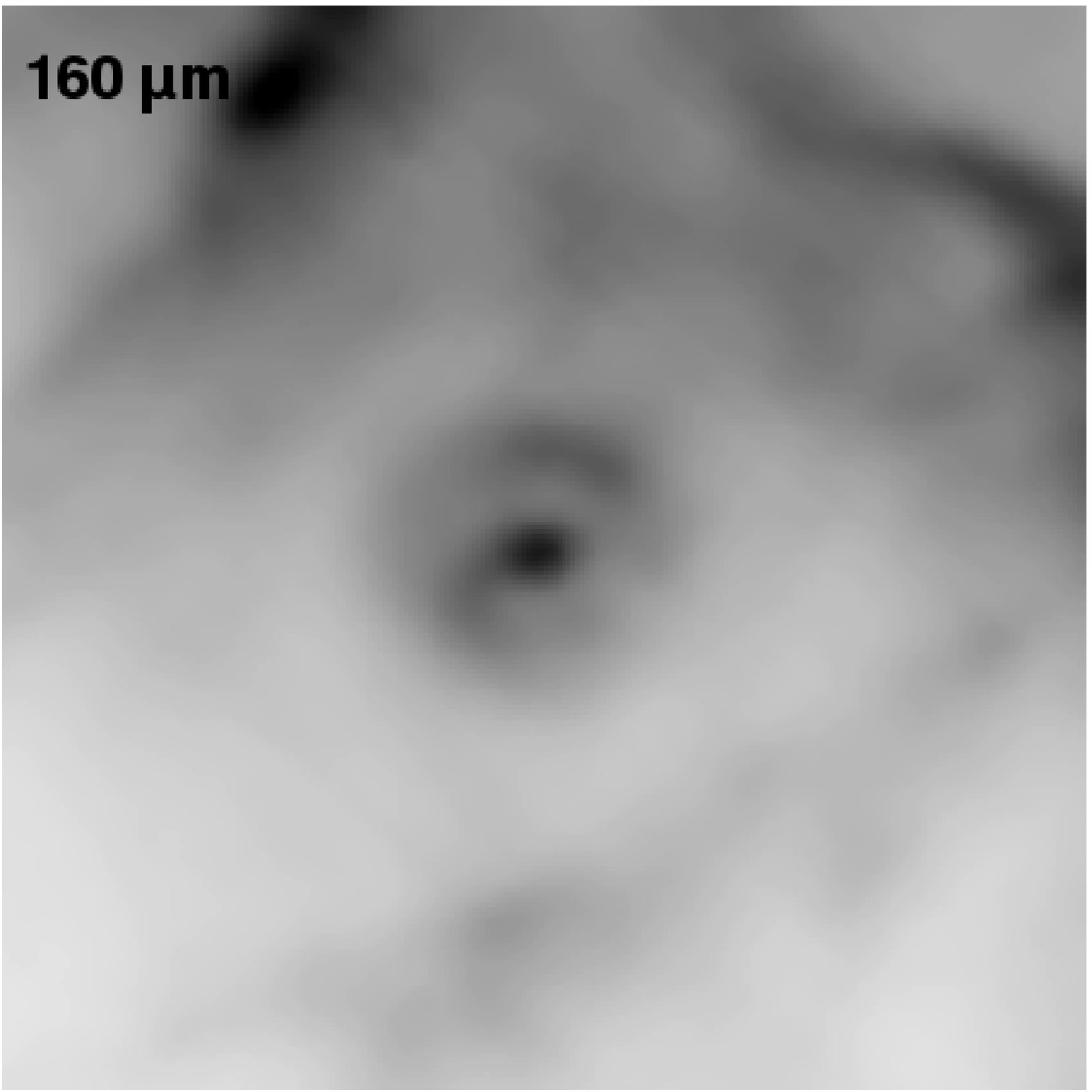}}\\
\resizebox{\hsize}{!}{\includegraphics*{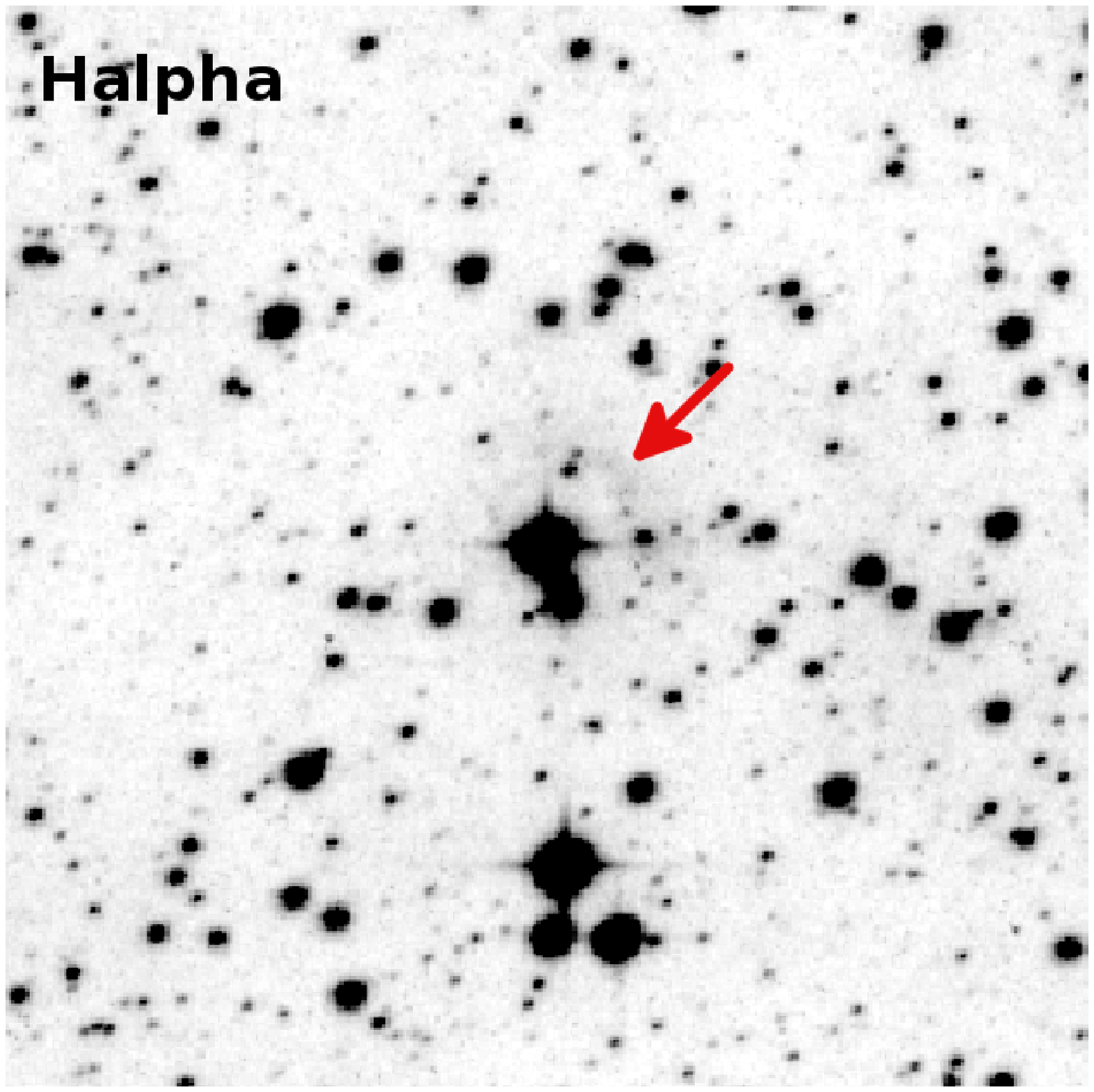}\includegraphics*{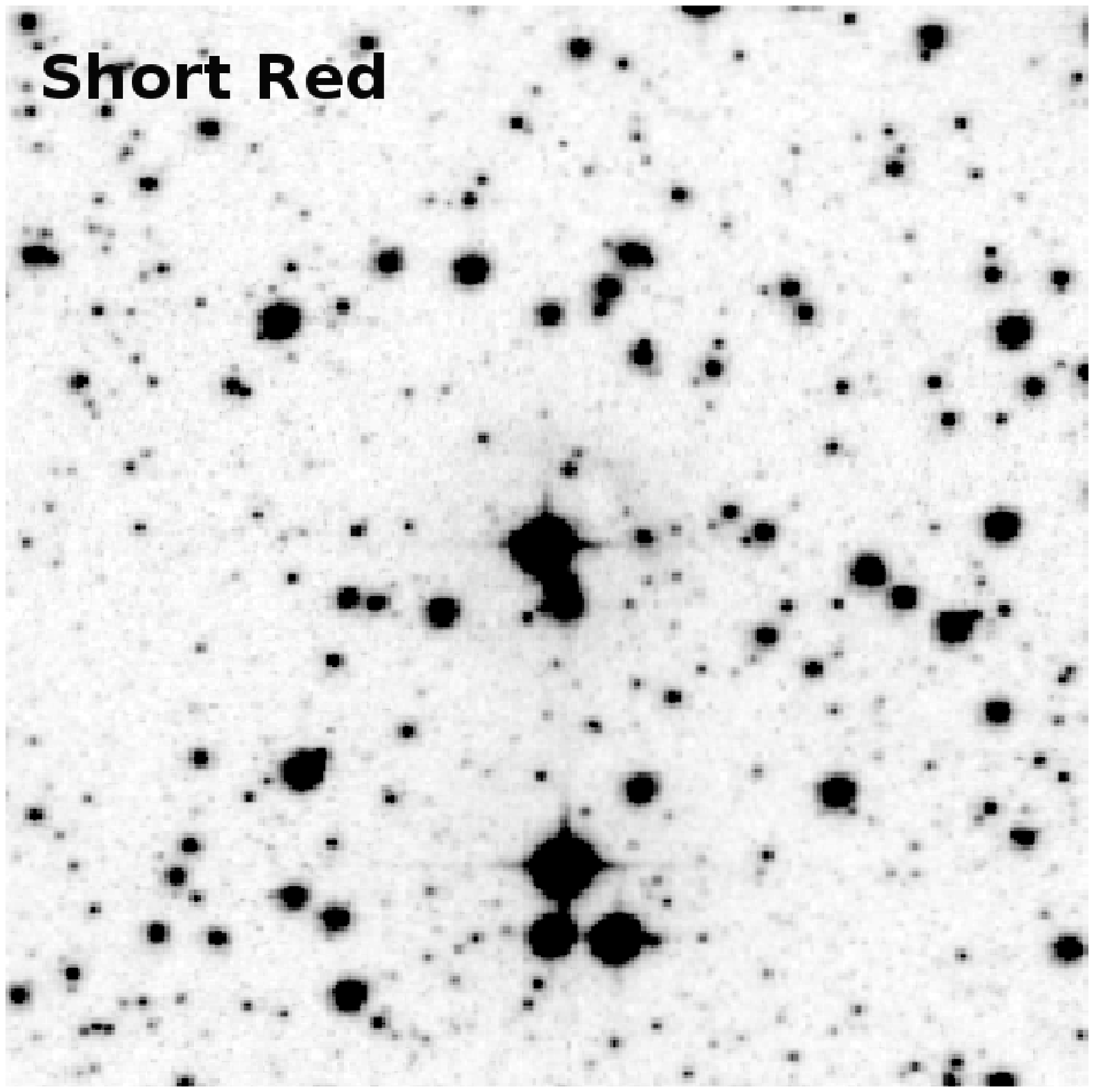}}
\caption{The 4\arcmin $\times$ 4\arcmin\ environment of \hen\ in the
70~$\mu$m and 160~$\mu$m {\it Herschel} PACS bands (top), and in the
H$\alpha$ and short red bands of the AAO/UKST SuperCOSMOS H$\alpha$
survey (bottom). North is up and East to the left. The arrow points to
a faint reflection arc NW of \hen .}
\label{ima4}
\end{figure}

Figures~\ref{rgb} and~\ref{ima4} illustrate the circumstellar
environment of \hen . A ring nebula is clearly observed in the
far-infrared images, with a diameter of approximately 50\arcsec.

The observations were carried out with the Photodetector Array Camera
and Spectrometer (PACS, Poglitsch et al. \cite{pog10}) onboard the
{\it Herschel} Space Observatory (Pilbratt et al. \cite{pil10}), in
the framework of the {\it Herschel} Infrared Galactic Plane survey
(Hi-GAL, Molinari et al. \cite{mol10}).  Hi-GAL is a key programme of
{\it Herschel}, mapping the inner part of the Galactic plane at 70 and
160 $\mu$m with PACS and 250, 350 and 500 $\mu$m with the Spectral and
Photometric Imaging Receiver (SPIRE, Griffin et al. \cite{gri10}).
The data are acquired in the PACS/SPIRE parallel mode by moving the
satellite at the speed of 60\arcsec s$^{-1}$.  This observing mode
results in image elongation along the scan direction.  In Hi-GAL two
orthogonal scans are secured.  This redundancy regularizes the PSF
which is roughly symmetric with FWHM of 10\arcsec\ and 13\arcsec\ at
70 and 160 $\mu$m respectively (Traficante et al. \cite{tra11}). The
observations, made immediately public for legacy, were retrieved from
the {\it Herschel} archive, pre-processed up to level 1 using the
Herschel Interactive Processing Environment (HIPE version 9; Ott
\cite{ott10}), and subsequently reduced and combined using
Scanamorphos (version 18; Roussel \cite{rou12}). Only PACS data were
considered, \hen\ being out of the SPIRE fields.

The dust nebula around \hen\ appears as a circular, clumpy ring,
extending from about 18\arcsec\ to 40\arcsec\ around the star with a
maximum at roughly 25\arcsec. At the distance of 4~kpc, this
corresponds to a radius of 0.5 pc, typical of LBV nebulae and
identical to the radius of the ring nebula around \wra .  The
morphology, round and clumpy, is similar to the morphology of the
inner ``Fried Egg'' nebula unveiled by Lagadec et al. (\cite{lag11})
using the Very Large Telescope Imager and Spectrometer for
mid-Infrared (VISIR). On the PACS images, the ``Fried Egg'' nebula is
unresolved and appears as the bright central spot observed at both 70
and 160 $\mu$m. The ring nebula is located in a cavity, best seen at
160 $\mu$m, 2\arcmin$-$3\arcmin\ ($\sim$3 pc) in diameter, which
delineate a bubble in the ambient interstellar medium, likely blown
out by the O star wind in a previous evolutionary phase, although a
remnant of an even older phase of mass-loss enhancement cannot be
excluded.

Attempts to detect an ionized gas nebula around \hen\ through
H$\alpha$ imaging failed, either from the ground (Le Bertre et
al. \cite{leb89}) or using the Hubble Space Telescope (Siodmiak et
al. \cite{sio08}). In Fig.~\ref{ima4} we show images from the AAO/UKST
SuperCOSMOS H$\alpha$ survey obtained in a narrow-band H$\alpha$+[NII]
and a broad-band Short Red filter, with bandwidths 6555-6625 \AA\ and
5900-6900 \AA\ respectively (Parker et al. \cite{par05}).  This survey
provides a 5-Rayleigh (i.e., $3 \times 10^{-17}$ erg cm$^{-2}$ s$^{-1}$
arcsec$^{-2}$) sensitivity with arcsecond spatial resolution.  A faint
arc is detected approximately 22\arcsec\ NW of the star, at a position
corresponding to the brightest part of the far-infrared ring. This arc
is equally seen in both filters and thus likely due to dust scattering
and not to H$\alpha$ emission. This confirms the absence of ionized
gas emission around \hen , in particular from the far-infrared dust
ring. For comparison, the ionized gas nebula around \wra\
(Hutsem\'ekers and Van Drom \cite{hut91}) is well detected on the
images of the SuperCOSMOS H$\alpha$ survey.

\section{Spectral energy distribution and nebular mass}
\label{sec:sed}

From the PACS images we measure the flux densities $F_{\nu}$ = 386
$\pm$3 Jy at 70 $\mu$m and 55 $\pm$3 Jy at 160 $\mu$m for the whole
nebula, using a 80\arcsec\ aperture and after subtraction of the
background emission estimated in the surrounding cavity.  The
contribution of the unresolved central source is 50 $\pm$2 Jy at 70
$\mu$m and 4.8 $\pm$0.5 Jy at 160 $\mu$m using a 12\arcsec\
aperture. In this case, we consider both the surrounding cavity and
the ring itself to estimate the background emission, the difference
being accounted for in the error budget. To account for the small
aperture, the flux measured at 70 $\mu$m has been increased by 10\%
and the flux measured at 160 $\mu$m by 25\% (Ibar et al. \cite{iba10};
we conservatively assume that the correction at 70 $\mu$m is of the
order of correction at 100 $\mu$m).

To build the spectral energy distribution (SED), we use the
ground-based near- and mid-infrared photometric measurements of
Le~Bertre et al. (\cite{leb89}) and Lagadec et
al. (\cite{lag11b}). These data only refer to the star and the inner
``Fried Egg'' nebula.  Other measurements were retrieved from the
NASA/IPAC Infrared Science Archive.  Only good quality data are
used. We consider the IRAS flux densities at 12, 25, and 60 $\mu$m,
the AKARI FIS measurements at 65 and 90 $\mu$m, and the MSXC6 data at
4.29, 4.35, 8.3, 12, 15, and 21 $\mu$m. The beam size of the IRAS and
AKARI observations is too large to resolve the different dust shells
so that these measurements refer to the whole nebula.

\begin{figure}[t]
\resizebox{\hsize}{!}{\includegraphics*{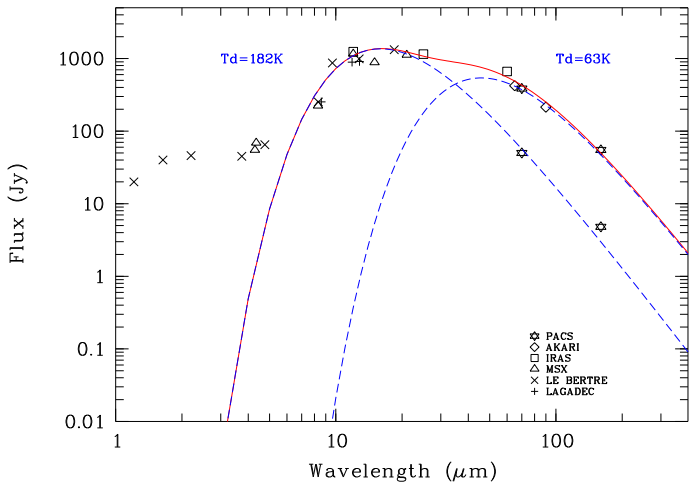}}
\caption{The spectral energy distribution of \hen . The different
symbols indicate photometric measurements from various sources. Error
bars are not showed for clarity but they are usually comparable to the
symbol size. At 70 $\mu$m and 160 $\mu$m, the two measurements refer
to the inner unresolved nebula and to the larger ring, both well
separated on the PACS images.  The SED is fitted with the sum
(red continuous curve) of two modified black-body curves with dust
temperatures $T_{\rm d}$ = 182~K and $T_{\rm d}$ = 63~K (dashed blue
curves).}
\label{fitbb}
\end{figure}

\begin{figure}[t]
\resizebox{\hsize}{!}{\includegraphics*{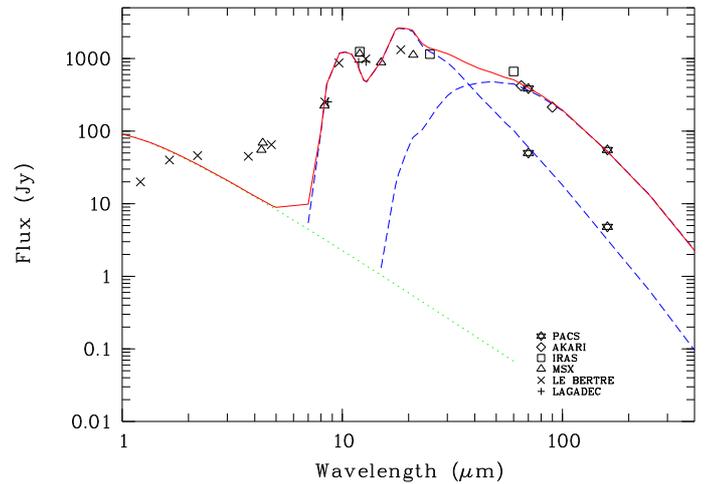}}
\caption{Same as Fig.\ref{fitbb}, but the spectral energy distribution
is fitted with the results of the 2-Dust radiative transfer code. The
total emission spectrum (red continuous curve) is shown as well as the
individual contributions of the inner ``Fried Egg'' nebula and the
outer ring (dashed blue curves). The stellar spectrum is
also shown (dotted green line).}
\label{fit2dust}
\end{figure}

All these data are plotted in Figs.~\ref{fitbb} and
~\ref{fit2dust}. They agree within the uncertainties and variability
limits. No color-correction has been applied, because it is small
(most often $< 10\%$) and difficult to evaluate given the complex SED
and the presence of strong silicate emission reported by Le Bertre
et al. (\cite{leb89}) on the basis of IRAS Low Resolution Spectra.

The total emission can be fitted as the sum of two modified black-body
curves $F_{\nu} \propto \nu ^{\beta} B_{\nu}(T_{\rm d})$. Since
silicates are clearly detected in the nebula around \hen\ (Le Bertre
et al. \cite{leb89}), we adopt $\beta = 2$. The contributions of the
inner ``Fried Egg'' nebula with a dust temperature $T_{\rm d}$ =182~K
and the larger ring nebula with $T_{\rm d}$ = 63~K are clearly
separated.

To further model the dust emission we use the two-dimensional
radiative transfer code 2-Dust (Ueta and Meixner \cite{uet03}). 2-Dust
is a versatile code which can be supplied with various grain size
distributions and optical properties as well as complex density
distributions. For 2-Dust, the inner radius of the dust shell
is an observable that can be measured from images so that the dust
temperature at that radius can be readily computed.

We assume that the nebula around \hen\ is spherically symmetric, which
is a good approximation of the overall geometry, and formed of two
separate shells, the first one being the ``Fried Egg'' nebula imaged
by Lagadec et al. (\cite{lag11}), with $r_{\rm in} = 0\farcs45$ and
$r_{\rm out} = 2\farcs25$, and the second one the large ring seen on
the PACS images, with $r_{\rm in} = 21\farcs5$ and $r_{\rm out} =
40\arcsec$. In the latter case, the exact value of $r_{\rm in}$ was
determined by comparing the PACS images to synthetic ones produced by
2-Dust and convolved with the PSF. In each shell we assume that the
dust density runs as $r^{-2}$. The addition of a third shell to better
account for the structure of the ``Fried Egg'' nebula appeared as an
unnecessary complication, only increasing the parameter space.  For
the stellar parameters, we use the luminosity $\log L / L_{\odot}$ =
5.7, the effective temperature $T_{\rm eff}$ = 8500 K, and the
distance $d$ = 4~kpc (Lagadec et al. \cite{lag11}). Since the dust in
\hen\ is dominated by silicates, we adopt a composition similar to the
one used for modeling \wra\ and other LBVs (Voors et al. \cite{voo00},
Vamvatira-Nakou et al., in prep.), i.e., pyroxenes with a 50/50 Fe
to Mg abundance (bulk density 3.2 g cm$^{-3}$). The optical constants
are taken from Dorschner et al. (\cite{dor95}) and extrapolated to a
constant refraction index in the far-ultraviolet. We assume a MRN size
distribution for the dust grains (Mathis et al. \cite{mat77}): $n(a)
\propto a^{-3.5}$ with $a_{\rm min} < a < a_{\rm max}$, $a$ denoting
the grain radius.  We did not attempt to fit the 5 $\mu$m emission
which is likely due to transiently heated dust grains and/or hot dust
very close to the star.  Adjustment to the data is mostly done by
tuning the optical depth, which controls the strength of the emission,
$a_{\rm max}$ which controls the 20 $\mu$m / 100 $\mu$m flux ratio in
the hotter shell, and the density ratio between the two shells.  The
two shells were considered simultaneously to reproduce the total
energy distribution as well as separately to measure their individual
contributions.  A good fit is obtained (Fig.~\ref{fit2dust}) except
for the 18 $\mu$m / 10 $\mu$m silicate emission band ratio which is
too high in the model, without incidence on our results.  Acceptable
values of $a_{\rm max}$ are found in the 1$-$3 $\mu$m range,
indicating the presence of large dust grains around \hen\ as in LBV
nebulae (Voors et al. \cite{voo00}).

Although there is some degeneracy in the parameter space, the derived
dust temperatures and masses are reasonably robust. From the 2-Dust
modeling, the temperature of the ``Fried Egg'' nebula is between 280
and 130 K, in agreement with Lagadec et al. (\cite{lag11}). For
the outer ring, the temperature is between 64 and 52 K. The dust mass
is $M_{\rm d}$ = $2.1 \times 10^{-3}$ $M_{\odot}$ for the ``Fried Egg''
nebula and $M_{\rm d}$ = 0.17 $M_{\odot}$ for the large ring.
Considering other acceptable models, we estimate the uncertainty of
the mass to be around 20$-$30\%.  $M_{\rm d}$ can also be derived
empirically using $M_{\rm d} = F_{\nu} \, d^{2} \, B_{\nu}^{-1} \,
K_{\nu}^{-1}$ where $K_{\nu}$ is the mass absorption coefficient
roughly independent of the grain radius (Hildebrand \cite{hil83}).
For the silicates of Dorschner et al. (\cite{dor95}), $K_{\nu} \simeq$
49 cm$^{2}$g$^{-1}$ at 70 $\mu$m. With $T_{\rm d}$ =182~K, we find
$M_{\rm d}$ = $1.4 \times 10^{-3}$ $M_{\odot}$ for the ``Fried Egg'' nebula,
and $M_{\rm d}$ = 0.12 $M_{\odot}$ for the larger ring nebula with
$T_{\rm d}$ = 63~K, in agreement with the 2-Dust results.

\section{\hen\ : a pre-LBV?}
\label{sec:discuss}

In Table~\ref{tab:compa} we compare \hen\ to \wra . The latter star is
a low-luminosity LBV surrounded by two massive dusty shells likely
ejected during a previous RSG phase as revealed by the C,N,O
abundances and in agreement with the evolutionary models of a star of
initial mass $\sim$ 40 $M_{\odot}$ with little rotation
(Vamvatira-Nakou et al., in prep.).  The two objects and their ejecta
appear very similar. In particular the outer ring nebula associated to
\hen\ has all the characteristics of a bona-fide LBV ring nebula,
except it is not photoionized.

Since LBV variations in the HR diagram are only reported when
observers catch them, one could imagine that \hen\ is a true LBV in
its cool phase, as supported by the similarity of its spectrum to the
1992 spectrum of HR Car (Le Bertre et al. \cite{leb93}), i.e., when HR
Car was itself in a cool phase (van Genderen et al. \cite{van97}).
However \hen\ has most probably not (yet) moved blueward.  Indeed if
the star had been as hot as \wra , it would have similarly
photoionized the ring nebula, stellar luminosities and nebular radii
being comparable.  Assuming comparable nebular densities, the
ionization/recombination time scale would be a few hundred years
(cf. Vamvatira-Nakou et al., in prep.) so that the ring nebula
around \hen\ should be glowing in H$\alpha$ as the \wra\ nebula, which
is not the case. Even a few decades spent in a hot phase would result
in a detectable H$\alpha$ nebula (Appendix \ref{app}).

\begin{table}[t]
\caption{Comparison of \wra\ and \hen }
\label{tab:compa}
\begin{tabular}{lcc}\hline\hline \\[-0.10in]
                                              &  \wra\                  &  \hen\    \\ 
\hline \\[-0.10in]
$\log L/L_{\odot}$                              & 5.7  $\pm$ 0.2          & 5.7 $\pm$ 0.1 \\
$T_{\rm eff}$  (10$^{3}$ K)                      & 30 $\leftrightarrow$ 9  & 8.5 $\pm$ 1.0 \\
$d$ (kpc)                                     & 6.0 $\pm$ 1.0           & 4.0 $\pm$ 0.5 \\
$r$ (inner shell) (pc)                        & 0.5                     & 0.04  \\
$r$ (outer shell) (pc)                        & 2.0                     & 0.5  \\
${\rm v}_{\rm exp}$ (km s$^{-1}$)                & 26                      & $\sim$ 30 \ \\
$t_{\rm kin}$ (inner shell) (10$^{4}$ yr)        & 1.9                     & 0.13  \\
$t_{\rm kin}$ (outer shell) (10$^{4}$ yr)        & 7.5                     & 1.6  \\
$M_{\rm d}$ (inner shell) (10$^{-2}$ $M_{\odot}$)  & 4.5 $\pm$ 0.5          & 0.21 $\pm$ 0.06 \\
$M_{\rm d}$ (outer shell) (10$^{-2}$ $M_{\odot}$)  & 5.0 $\pm$ 2.0          & 17 $\pm$ 5 \\
\hline\\[-0.2cm]
\end{tabular}\\
\tiny{Source: Sterken et al. \cite{ste08}; Vamvatira-Nakou et
al., in prep.; Lagadec et al. \cite{lag11}; this work. The
expansion velocity of the \hen\ nebula is assumed similar to the one
measured for the shell around yellow hypergiant \irc , i.e.,
${\rm v}_{\rm exp} \simeq$ 25-37 km s$^{-1}$ (Castro-Carrizo et
al. \cite{cas07}). The inner shell of \hen\ refers to the whole
``Fried Egg'' nebula and the outer shell to the large ring nebula.}
\end{table}

According to the kinematical age of the nebulae, \hen\ looks younger
than \wra , with the inner shell possibly still in formation. \hen\ is
most likely in a pre-LBV evolutionary stage.  The kinematical age of
the ring nebula around \hen , $1.6 \times 10^{4}$ yr, indicates that the star
has already spend a significant fraction of its time as a post-red
supergiant star (cf. the models for a 40 $M_{\odot}$ star by Ekstr\"om
et al. \cite{eks12}). This also suggests that the ring nebula should
have been ejected close to the beginning of the RSG phase, possibly
due to the dynamical instability mechanism proposed by Stothers and
Chin (\cite{sto96}). If we adopt a gas to dust ratio of 40 as measured
for the \wra\ nebula (Vamvatira-Nakou et al., in prep.),
\hen\ has lost $\sim$ 7 $M_{\odot}$ of gas and dust since the
beginning of the RSG phase. Interestingly enough, \wra\ has ejected
two shells of comparable mass, while for \hen , most of the mass is
concentrated in the outer ring.

\section{Conclusions}
\label{sec:conclu}

We have found a large dust ring nebula around the yellow hypergiant
\hen . With 1 pc in diameter and a total gas mass estimated to 7
$M_{\odot}$, this nebula is comparable to the nebulae observed around
LBVs.  In particular, \hen\ appears very similar to the low-luminosity
(loL; $\log L/L_{\odot} \lesssim 5.8$) LBV \wra . The fact that the
nebula is not seen in H$\alpha$ indicates that \hen\ has not yet moved
to a hotter phase, and is still in a pre-LBV stage.  \hen\ might be
just ready to cross the yellow void and become a hot LBV, a scenario
also suggested for \irc\ (Humphreys et al. \cite{hum02}).  Our
observations strongly support the evolutionary path: RSG $\rightarrow$
YHG $\rightarrow$ loL-LBV, providing direct evidence that massive LBV
nebulae can be ejected during the RSG phase.

\begin{acknowledgements}
DH, NLJC and CVN acknowledge support from the Belgian Federal Science
Policy Office via the PRODEX Programme of ESA.  PACS has been
developed by a consortium of institutes led by MPE (Germany) and
including UVIE (Austria); KU Leuven, CSL, IMEC (Belgium); CEA, LAM
(France); MPIA (Germany); INAF-IFSI/OAA/OAP/OAT, LENS, SISSA (Italy);
IAC (Spain).  This development has been supported by the funding
agencies BMVIT (Austria), ESA-PRODEX (Belgium), CEA/CNES (France), DLR
(Germany), ASI/INAF (Italy), and CICYT/MCYT (Spain).  This research
has made use of the NASA/IPAC Infrared Science Archive, which is
operated by the Jet Propulsion Laboratory, California Institute of
Technology.
\end{acknowledgements}

\newpage

\appendix

\section{Detection limit of the nebula in H$\alpha$}
\label{app}

We can roughly estimate the time needed in the hot phase to produce a
detectable H$\alpha$ nebula.  Given its sensitivity
(Sect.~\ref{sec:imaging}), the SuperCOSMOS survey can detect
$F(\rm{H}\alpha) > \, 6 \times 10^{-14}$ erg cm$^{-2}$ s$^{-1}$ from a
(homogeneous) ionized nebula 25\arcsec\ in radius. Within a time $t$,
a star emitting $Q_0$ ionizing photon s$^{-1}$ ionizes $Q_0 \, t$
hydrogen atoms. The ionized gas will recombine and emits $F({\rm
H}\alpha) = (1/4\pi) \, n_e \, n_p \, h \nu \, \alpha^{\rm eff}_{{\rm
H}\alpha} \, V \, d^{-2}$ where $n_e$ and $n_p$ are the electron and
proton densities, $\alpha^{\rm eff}_{{\rm H}\alpha}$ the effective
recombination coefficient, $\nu$ the frequency of H$\alpha$, $h$ the
Planck constant, $V$ the emitting volume and $d$ the distance to the
nebula. Making $n_e \, n_p = (Q_0 \, t / V)^{2}$ we can estimate the
time that the star must emit $Q_0$ ionizing photon s$^{-1}$ to make
its surrounding nebula glowing in H$\alpha$ above the detection limit
of the SuperCOSMOS survey. With $Q_0$ = 10$^{47}$ photon s$^{-1}$ as
found for \wra\ (Vamvatira-Nakou et al., in prep.), we find $t
\sim$ 20 years.


\begin{thebibliography}{}
  \bibitem[2007] {cas07} Castro-Carrizo, A., Quintana-Lacaci, G., Bujarrabal, V., Neri, R., \& Alcolea, J. \ 2007, \aap, 465, 457
  \bibitem[1995] {dor95} Dorschner, J., Begemann, B., Henning, T., Jaeger, C., \& Mutschke, H.\ 1995, \aap, 300, 503
  \bibitem[2012] {eks12} Ekstr\"om, S., Georgy, C., Eggenberger, P., et al. \ 2012, \aap, 537, A146
  \bibitem[2010] {gri10} Griffin, M. J., Abergel, A., Abreu, A., et al. \ 2010, \aap, 518, L3
  \bibitem[1976] {hen76} Henize, K.G. \ 1976, \apjs, 30, 491
  \bibitem[1983] {hil83} Hildebrand, R.~H.\ 1983, \qjras, 24, 267
  \bibitem[1994] {hum94} Humphreys, R.M. \& Davidson, K.\ 1994,  \pasp, 106, 1025
  \bibitem[2002] {hum02} Humphreys, R.M. \& Davidson, K., Smith, N. \ 2002,  \apj, 124, 1026
  \bibitem[1991] {hut91} Hutsem\'ekers, D. \& Van Drom, E.\ 1991, \aap, 251, 620 
  \bibitem[1994] {hut94} Hutsem\'ekers, D.\ 1994,  \aap, 281, L81
  \bibitem[2010] {iba10} Ibar, E., Ivison, R.J., Cava, A., et al. \ 2010, \mnras, 409, 38
  \bibitem[2001] {lam01} Lamers, H. J. G. L. M., Nota, A., Panagia, N., Smith, L. J. \& Langer, N.\ 2001, \apj, 551, 764
  \bibitem[2011] {lag11} Lagadec, E., Zijlstra, A.A., Oudmaijer, R.D., et al. \ 2011, \aap, 534, L10
  \bibitem[2011b]{lag11b} Lagadec, E., Verhoelst, T., Mekarnia, D., et al. \ 2011, \mnras, 417, 32 
  \bibitem[1989] {leb89} Le Bertre, T., Epchtein, N., Gouiffes, C., Heydari-Malayeri, M., \& Perrier, C. \ 1989, \aap, 225, 417
  \bibitem[1993] {leb93} Le Bertre, T., Lequeux, J. \ 1993, \aap, 274, 909
  \bibitem[1995] {not95} Nota A., Livio M., Clampin M. \& Schulte-Ladbeck R.\ 1995,  \apj, 448, 788
  \bibitem[2010] {mae10} Maeder, A. \& Meynet, G.\ 2010, NewAR, 54, 32
  \bibitem[1977] {mat77} Mathis, J.~S., Rumpl, W., \& Nordsieck, K.~H.\ 1977, \apj, 217, 425
  \bibitem[2010] {mol10} Molinari, S., Swinyard, B., Bally, J., et al. \ 2010, \pasp, 122, 314 
  \bibitem[2009] {oud09} Oudmaijer, R.D., Davies, B., de Wit, W.-J.,  \& Patel, M. \ 2009, ASP Conference Series, 412, 17
  \bibitem[2010] {ott10} Ott, S.\ 2010,  ASP Conference Series, 434, 139
  \bibitem[2005] {par05} Parker, Q.A., Phillipps, S., Pierce, M.J., et al. 2005, \mnras, 362, 689
  \bibitem[2010] {pil10} Pilbratt, G.L., Riedinger, J.R., Passvogel, T., et al.\ 2010,  \aap, 518, 1
  \bibitem[2010] {pog10} Poglitsch, A., Waelkens, C., Geis, N., et al.\ 2010,  \aap, 518, 2
  \bibitem[2008] {ste08} Sterken, C., van Genderen, A.~M., Plummer, A., \& Jones, A.~F.\ 2008, \aap, 484, 463
  \bibitem[1996] {sto96} Stothers, R.,B., Chin, C.-W., \ 1996, \apj, 468, 842
  \bibitem[2012] {rou12} Roussel, H. \ 2012, arXiv:1205.2576 
  \bibitem[2008] {sio08} Siodmiak, N., Meixner, M., Ueta, T., Sugerman, B. E. K., Van de Steene, G. C.,  \& Szczerba, R. \ 2008, \apj, 677, 382
  \bibitem[2011] {tra11} Traficante, A., Calzoletti, L., Veneziani, M., et al. \ 2011, \mnras, 416, 2932
  \bibitem[2003] {uet03} Ueta, T., \& Meixner, M.\ 2003, \apj, 586, 1338
  \bibitem[1997] {van97} van Genderen, A.M., de Groot, M., \&  Sterken, C. 1997, \aaps, 124, 51  
  \bibitem[2000] {voo00} Voors, R. H. M., Waters, L. B. F. M. , de Koter, A., et al.\ 2000, \aap, 356, 501
  \bibitem[1997] {wat97} Waters, L. B. F. M., Morris, P. W., Voors, R. H. M. \& Lamers, H. J. G. L. M., 1997, ASP Conference Series, 120, 326
\end{thebibliography}
\end{document}